\documentclass{article}
\usepackage{verbatim}
\usepackage{moreverb}
\usepackage{url}
\usepackage{amsmath}
\author{Vadim Zaliva, lord@accelkey.com}
\date{\today}
\title{{\sl AccelKey} Selection Method for Mobile Devices}
\usepackage{graphicx}
\usepackage{subfig}
\usepackage[colorlinks=true,bookmarks=true,pdfauthor={Vadim Zaliva, lord@crocodile.org},
            pdftitle={AccelKey: List Selection Method for Mobile Devices},
            pdftex]{hyperref}

\begin{document}

\maketitle

\begin{abstract}
Portable Electronic Devices usually utilize a small screen with limited viewing area and a keyboard with a limited number of keys. This makes it difficult to perform quick searches in data arrays containing more than dozen items such an address book or song list. In this article we present a new data selection method which allows the user to quickly select an entry from a list using 4-way navigation device such as joystick, trackball or 4-way key pad. This method allows for quick navigation using just one hand, without looking at the screen.
\end{abstract}

\tableofcontents

\section{Introduction}

Selecting information from lists on a mobile device is a common task, but the process can be quite cumbersome. A typical example of this is finding a name in an address book to make a phone call. If the list contains more than a dozen entries, scrolling through it to reach a name becomes impractical. Not all mobile devices have a keyboard. Navigating using virtual keyboard displayed on a touch screen usually requires use of a stylus. Taking out a stylus adds additional time, and the use of a stylus itself is relatively slow. One can use a regular phone keyboard to search for a name, but most selection methods offered by modern mobile phones are not very efficient. Also, some modern mobile phones and PDAs are using sliding design, where the keyboard is normally hidden and has to be slid out to be used (e.g. Siemens SX66). Sliding out or opening keyboard takes additional time.

\begin{figure}[htp]
\centering
\subfloat[Blackberry PEARL]{
  \includegraphics[width=1.5in]{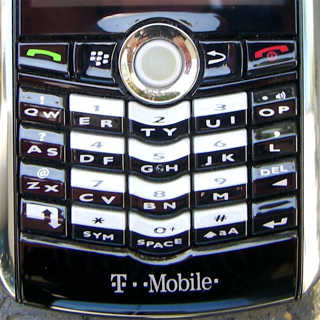}
  \label{fig:BlackberryPEARL}
}
\subfloat[I-Mate SP3i]{
  \includegraphics[width=1.5in]{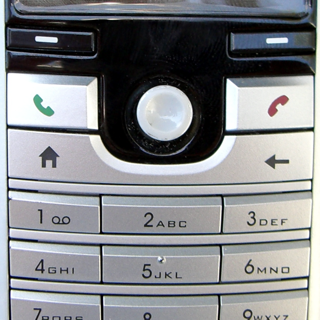}
  \label{fig:IMateSP3i}
}
\subfloat[Nokia 2865]{
  \includegraphics[width=1.5in]{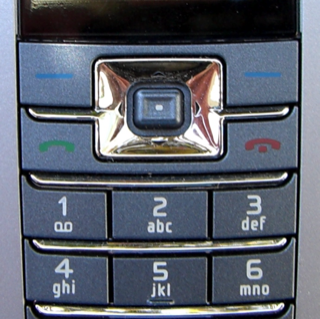}
  \label{fig:Nokia2865}
}
\subfloat[iPAQ h6315]{
  \includegraphics[width=1.5in]{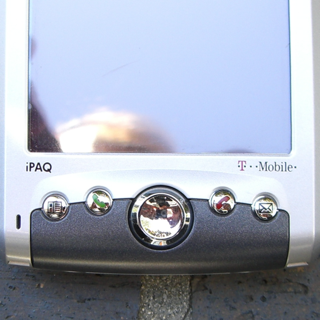}
  \label{fig:iPAQh6315}
}
\subfloat[Motorola RAZR]{
  \includegraphics[width=1.5in]{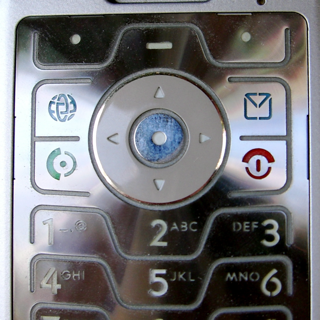}
  \label{fig:MotorolaRAZR}
}
\subfloat[Sanyo MM-8300]{
  \includegraphics[width=1.5in]{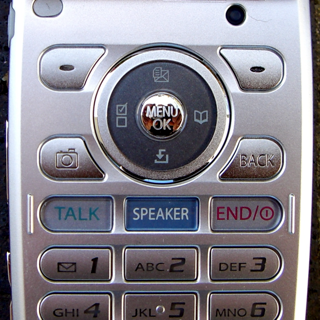}
  \label{fig:SanyoMM8300}
}
\subfloat[Motorola V710]{
  \includegraphics[width=1.5in]{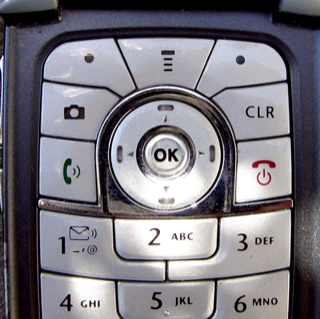}
  \label{fig:MotorolaV710}
}
\subfloat[SonyEricsson Z520a]{
  \includegraphics[width=1.5in]{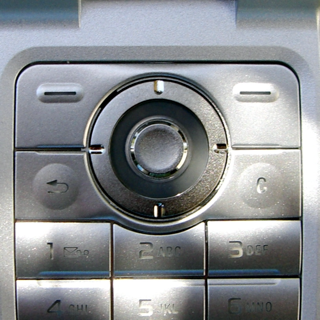}
  \label{fig:SonyEricssonZ520a}
}
\caption{Examples of Mobile Phones Input Devices}
\label{fig:phones}
\end{figure}

In this paper, we would like to present a novel approach to this problem. Instead of phone keyboards, a user would use a joystick, a trackball or 4-way key pad. Most modern mobile phones are equipped with one of these navigation devices. In Figure~\ref{fig:phones}, you can see some actual mobile phones which are suitable for use with the {\sl AccelKey} input method. The device shown in \ref{fig:BlackberryPEARL} uses a trackball, \ref{fig:IMateSP3i}, \ref{fig:Nokia2865},\ref{fig:SonyEricssonZ520a} use joysticks, and \ref{fig:iPAQh6315}, \ref{fig:MotorolaRAZR}, \ref{fig:SanyoMM8300},\ref{fig:MotorolaV710} each use a four-button pad.

The presented method is a \textit{selection method}, not an \textit{input method}. It does not allow the user to enter arbitrary data.  Rather, it simply selects an entry from the existing dataset. \cite[MacKenzie‌, Soukoreff (2002)]{textentry} provides a good overview of various text input methods.‌

The presented method is suitable for one-handed, "eyes-free" operation\footnote{Using \emph{focus of attention}  notation\cite{textentry}, this method could be classified as single FOA task.}.

\section{{\sl AccelKey} Algorithm}

In this section, we will give a formal description of the algorithm. Please see Appendix~\ref{fast4hs} for a sample implementation of the algorithm in a Haskell programming language.

A user is presented with a list of entries. His goal is to select one of the entries from the list using a 4-way navigation device. The device is capable of generating at least five events triggered by user actions. Directional events \emph{up}, \emph{down}, \emph{left}, and \emph{right} are generated when the joystick is tilted or the trackball is rotated in the appropriate direction. An additional \emph{select} event is generated when the user pushes the joystick or trackball down, or presses a dedicated button. Additional events, such as \emph{backspace} or \emph{reset} could be assigned to to further keys to improve user experience, but are not strictly necessary for this algorithm. They are covered in Section~\ref{extrasig}.

Each directional event is associated with a group of letters from the alphabet from which elements of the list are composed.  We will call such assignments a ``Layout''. Issues related to layout selection will be covered in the Section~\ref{layoutsel}. For now, let us just assume the letter assignment is non-overlapping --- i.e., each letter is associated only with one of four directional events. Not all characters of the alphabet have to be associated with directional events. Characters which are not associated with events (e.g punctuation marks and whitespace) are deemed insignificant and ignored by the algorithm.

The algorithm could be in one or two modes: \emph{selection} or
\emph{scrolling}. The initial state is \emph{selection}.

\subsection{\emph{Selection} Mode}

When in this mode, the algorithm maintains a sequence of directional events generated by the user. We will call this sequence a \emph{prefix}. Initially, the prefix is empty. Each new directional event generated by the user is appended to the end of the prefix. The list is then filtered by excluding all entries except ones matching the current prefix. The entry is matching the prefix if, and only if in the current layout each of its letters belongs to the group associated with the prefix event with the same index.\footnote{Case-sensitive or  case-insensitive check could be performed, depending on the nature  of the data presented in the list.}

At any moment the user can chose to generate a \emph{select} event. When this happens, if the current, filtered list contains only one element the selection process is completed. If the list contains more than one element, we switch to the \emph{scrolling} mode described in Section~\ref{sst} to narrow down the selection further.

This mode can be likened to presenting a user with a keyboard with 4 keys: the four keys have groups of letters on them, and there is an additional \emph{select} key. The user types his search term using this keboard, and the algorithm performs disambiguation using list content as a dictionary. In some respects this is similar to the Tegic T9\textsuperscript{\textregistered} input method.

\subsection{\emph{Scrolling} Mode}
\label{sst}

Upon entering this mode one of the list elements is selected as 
current.\footnote{Initial selection of a current element is not as simple as it seems.  Though the first entry might seem to be an obvious choice, selecting the element in the middle of the list instead would minimize the average required number or keystrokes. If the list is too long to fit on the screen, the sort order becomes important, because the user should be able to decide in which direction to scroll   based on information he or she sees on the screen. For this, the list must be sorted. The sorting order and criteria are not important for the algorithm, as long as it obvious to the user in which direction to scroll from the current element to reach the element he is looking for. If the first element is selected   initially, the sorting requirement can be lifted since the only direction the user can scroll is down.} Using \emph{up} and \emph{down} events, the user then navigates towards the entry he is looking for and selects it using the \emph{select} event. 
Directional events \emph{up} and \emph{down} move the selection pointer, selecting previous or next list elements as current.  Any events of \emph{left} or \emph{right} will bring the algorithm back to \emph{selection} state, using the last used prefix.  Finally, \emph{select} will indicate that the current element is the ultimate user selection.

\subsection{Comparison to Other Methods}

In this section, we will compare {\sl AccelKey} to a few other selection algorithms.  The algorithms we have chosen to evaluate are:\paragraph*{Scroll} - This is the simplest possible algorithm. A user is presented with the list and a cursor. The cursor initially points to the first element. Using up and down arrows, a user can select the desired list element. \paragraph*{Multi-Tap First} - This is the only selection algorithm which is implemented in some mobile phones (for example Motorola V600) besides \emph{Scroll} algorithm. A user can select the first letter of the name using the \emph{multi-tap} method\cite{multitap}. Once the first letter is selected, the cursor is positioned to the first entry starting with this letter and then, user can scroll down to select a desired name using \emph{scroll} method.\paragraph*{Multi-Tap Match} - This is a more advanced method, implemented in some phones (e.g. Sony Ericsson K750). A user can type part of the name using multi-tap input method. As he types the cursor moves to the first name that matches the typed prefix. At any moment the user can switch to the \emph{scroll} method to continue selection. In our evaluation, we implemented the best possible strategy, deciding when it was the best time to switch from typing to scrolling. We suspect most users might not perform as well as our program, but we will use these numbers for an evaluation, as a best case scenario.  To compare the efficiency of these algorithms we took several datasets and ran all algorithms on each of them. The measure of efficiency was an average number of \textit{events} (user actions such as key presses or joystick movements) required to select any name from the dataset. The evaluation was performed using a program which is included in the Appendix~\ref{src}. The {\sl AccelKey} algorithm used a \emph{QWERTY} layout (see Section~\ref{layoutsel}). The results are shown in the Table~\ref{tb:mcmp}.

\begin{table}[h]
\begin{center}  
\begin{tabular}{|l|l|l|l|l|}
\hline
&\multicolumn{4}{c}{{\bf Method}}\vline\\
\hline
{\bf Dataset} & {\sl AccelKey} & {\sl Scroll} & {\sl Multi-Tap} & {\sl Multi-Tap First}\\
\hline
Writers & 4.08 & 47.5 & 4.66 & 4.67 \\
\hline
Representatives & 5.16 & 196.5 & 6.21 & 6.22 \\ 
\hline
Graduates & 7.11 & 684 & 7.58 & 7.58 \\ 
\hline
\end{tabular}
\end{center}
\caption{Selection Methods Efficiency Evaluation (average number of events per entry)}
\label{tb:mcmp}
\end{table}

Following datasets were used:

\paragraph*{Writers} This dataset consists of the ``100 Best Writers of the 20th Century''\cite{100best}. It contains 96 entries. Only last names were used.\paragraph*{Representatives} The dataset is a list of names of members of U.S. House of Representatives.\cite{repr} It contains 394 entries. Only last names were used.

\paragraph*{Graduates} The dataset is a list of names of students to receive degrees at University of New Brunswick in May 2004.\cite{grads}  The list contains 1369 entries. Only last names were used.

As we can see, on all three datasets {\sl AccelKey} algorithm requires fewer keystrokes than others. Also, since it is usually can be operated without removing one's finger from the joystick, there is no time cost of moving fingers between keys.

\subsection{Layout Evaluation}
\label{layoutsel}

Selected layout (the association between letters and directional events) can substantially affect the performance of {\sl AccelKey}. There are a few approaches to optimize layout:\paragraph*{Mnemonically} - to make the layout easy to remember. One obvious layout is alphabetical. Another possible layout is based on the QWERTY keyboard\cite{qwerty} layout. Experienced computer users may find it easier to use than alphabetical, since usually one can remember at least the general location of keys on his or her keyboard.\paragraph*{Statistically} - to minimize the number of events required to select any list element. Here, the layout would depend on a corpus: letter frequencies, probabilities of one letter following another, etc. Layout optimization algorithms is one of the subject of our future research.  Using source code from Appendix~\ref{src} we have evaluated two layouts: 

\paragraph*{\emph{ABC} Layout} splits the Latin alphabet in its traditional order into 4 groups: {\sl[A,B,C,D,E,F,G]}  (\emph{up}), {\sl[H,I,J,K,L,M,N]} (\emph{left}), {\sl[O,P,Q,R,S,T,U]} (\emph{right}), {\sl [V,W,X,Y,Z]} (\emph{down}).

\paragraph*{\emph{QWERTY} Layout} rougly emulates key positions on a personal computer keyboard (QWERTY layout\cite{qwerty}, US version), arranging letters into four groups: {\sl[Q,W,E,R,T,Y,U,I,O,P]} (\emph{up}), {\sl[A,S,D,F,G]} (\emph{left}), {\sl[H,J,K,L]} (\emph{right}), {\sl[Z,X,C,V,B,N,M]}  (\emph{down}).

Using datasets used in previous section, the average number of events to select any dataset element using a given layout is shown in Table~\ref{tb:lcmp}.

\begin{table}[h]
\begin{center}  
\begin{tabular}{|l|l|l|}
\hline
&\multicolumn{2}{c}{{\bf Layout}}\vline\\
\hline
{\bf Dataset} & {\sl QWERTY} & {\sl ABC} \\
\hline
Writers & 4.08 & 4.07 \\
\hline
Representatives & 5.16 & 5.24 \\ 
\hline
Graduates & 7.11 & 6.76 \\ 
\hline
\end{tabular}
\end{center}
\caption{Layout Efficiency (average number of events per entry)}
\label{tb:lcmp}
\end{table}

As we can see, on the \emph{Writers} and \emph{Graduates} datasets, an \emph{ABC} layout has shown to give better results, while \emph{QWERTY} layout fared better with the \emph{Representatives} dataset.

\section{Algorithm Extensions}

\subsection{Additional Events}
\label{extrasig}
For user convenience, we can introduce some additional events. These events are not essential to the core algorithm operation, and their implementation is optional. \emph{Backspace} - eliminates effects of the previous event and brings the algorithm to the previous state. \emph{Reset} - resets the algorithm to the initial state, which is \emph{scrolling} mode, with an empty prefix.

\subsection{Using Keyboard or KeyPad}

On devices equipped with a keyboard in addtition to a 4-directional input device, AccelKey selection algorithm could be combined with more traditional input methods.

For example, on mobile phones equipped with a standard phone numeric keyboard, numbered keys could be generating events which are treated as an additional nine events using special layout: {\sl[A,B,C]}  (\emph{Key ``2''}), {\sl[D,E,F]} (\emph{Key ``3''}), {\sl[G,H,J]} (\emph{Key ``4''}), {\sl [J,K,L]} (\emph{Key ``5''}), {\sl [M,N,O]} (\emph{Key ``6''}), {\sl [P,Q,R,S]} (\emph{Key ``7''}), {\sl [T,U,V]} (\emph{Key ``8''}), {\sl [W,X,Y,Z]} (\emph{Key ``9''}). Thus, the user could mix directional events from a joystick and keypad events during the selection process.

On devices equipped with a \emph{QWERTY} (or similar) keyboard, users should be able to combine AccelKey directional events with keypreses from a keyboard during the selection process. For example, tilting a joystick upwards and then pressing the ``z'' key on a keyboard would select all entries which start from any letter associated with the \emph{up} event in the current layout, followed by the letter ``z''.

\subsection{Using trackball}
It is possible to use AccelKey with devices equipped with a trackball (for example, BlackBerry Pearl). Unlike a joystick which generates descrete directional events, a trackball generates events containing the following values:

\begin{description}
\item[$\Delta x$] Magnitude of navigational motion: negative for a move left and postive for a move right.
\item[$\Delta y$] Magnitude of navigational motion: negative for an downwards move, and positive for a upwards.
\end{description}

Mapping between these values and directional events used by AccelKey algorithm could be expessed as:

\begin{equation*}
event(\Delta x, \Delta y)=\begin{cases} 
\emph{right} & \text{, if $\lvert \Delta x \rvert > \lvert \Delta y \rvert $ and $\Delta x > 0$},\\ 
\emph{left} & \text{, if $\lvert \Delta x \rvert > \lvert \Delta y \rvert $ and $\Delta x \leq 0$},\\ 
\emph{up} & \text{, if $\lvert \Delta x \rvert \leq \lvert \Delta y \rvert $ and $\Delta y > 0$},\\ 
\emph{down} & \text{, if $\lvert \Delta x \rvert \leq \lvert \Delta y \rvert $ and $\Delta x \leq 0$},\\ 
\end{cases} 
\end{equation*}

\begin{figure}[htp]
\centering
\includegraphics[width=4in]{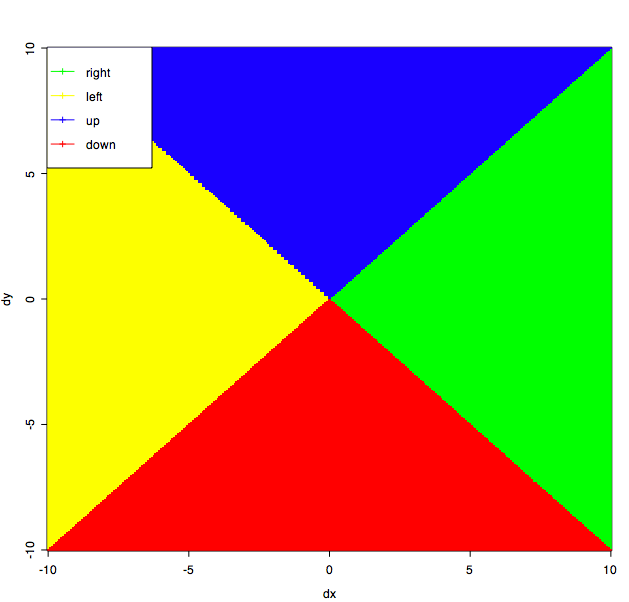}
\caption{Mapping trackball movement offsets to directional events}
\label{fig:joydir}
\end{figure}

This mapping for $\Delta x$ and $\Delta x$ in $\left[-10,10\right]$ interval is shown in Figure~\ref{fig:joydir}. 

Some trackball devices are very sensitive and report even very slight movements, which makes it necessary to implement \emph{jitter elimination}. The simplest form is to ignore trackball events with movement amplitude less than a certain value. In other words, ignore trackball events if $\sqrt{{\Delta x}^2 +{\Delta y}^2}$ is less than some pre-defined small value.

\subsection{Multiple Word Selection}
The algorithm could also be extended to handle lists with multiple words per entry. One example of such a list is an address book, where for each entry you can commonly have a combination of first, last, and middle names.

The most obvious approach here is to simultaneously match the \textit{prefix} to the beginning of the first, last and middle names, thus allowing a user to select using any of these fields. For example, to select {\sl Jorge Luis Borges} the user can enter {\sl Jorge} or {\sl Luis} or {\sl Borges}\footnote{This could be extended to match as many words as needed, for example to use full name like {\sl Jorge Francisco Isidoro Luis Borges Acevedo}.}.

\begin{figure}[htp]
\centering
\includegraphics[width=2in]{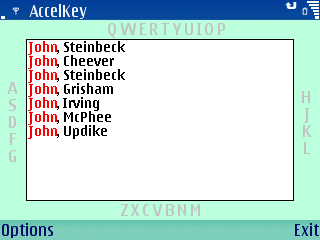}
\caption{Example of an address book selection using common first name}
\label{fig:johns}
\end{figure}

Though a multiple word match approach is useful, it could be further improved upon. For example, selecting a prefix or matching a common first name presents us with a fairly long list of choices which cannot be narrowed down further without switching to a selection mode. One example of such a case is shown in Figure~\ref{fig:johns}.

To solve this, we can allow for prefix matching to span boundaries of adjacent words. Thus, in the example shown in Figure~\ref{fig:johns}, the user should be able to generate events matching letters: ``j'', ``h'', ``o'', ``n'', ``u'' to narrow the list down to single entry: ``John Updike''. Of course, prefix match could span more than two words, as long as all of them match sequentially (for example ``Arthur,C.,Clarke'' could be matched as ``a'', ``r'', ``t'', ``h'', ``u'', ``r'', ``c'', ``c'').

In some applications, we might want to allow matching to ``wrap'' at the end of the last word in a sequence and to proceed from the beginning of first word. For example, this would allow us to match ``Smith, John''  as ``j'', ``h'', ``o'', ``n'', ``s''. If matching is always to stop at the last word in a group, there is a no way to disambiguate ``Smith, John'' from ``Blake, John'' after the first name was matched.

\section{Implementation}

\begin{figure}[htp]
\centering
\includegraphics{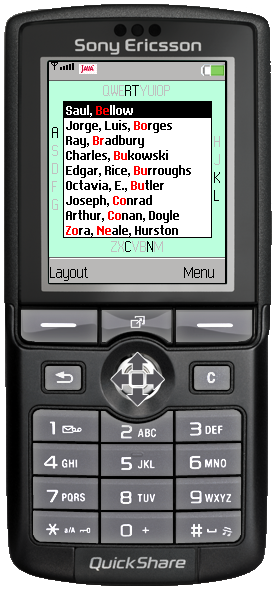}
\caption{AccelKey running on mobile phone}
\label{fig:Fast4SEFull}
\end{figure}

The algorithm was implemented as a Java Applet, BlackBerry application, J2ME MIDlet and SymbianOS application, and has been tested on several mobile phones. The latest demo version can be downloaded from the following URL: \url{http://www.accelkey.com/}

The {\sl AccelKey} application is shown running on a mobile phone in Figure~\ref{fig:Fast4SEFull}.

\begin{figure}[htp]
\centering
\includegraphics{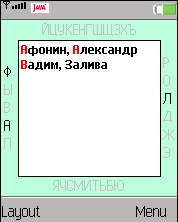}
\caption{{\sl AccelKey} with Cyrillic Alphabet}
\label{fig:screen_cyr}
\end{figure}

Let us examine the screen in more detail. In the middle of the screen is the list of names from which we will be selecting. They are displayed in \textit{First, Middle, Last Name} format. As a user sends more events using the joystick, the content of this list is narrowed down to contain only names matching the current \emph{prefix}.  In each name, the letters matching the current \emph{prefix} are shown in red. During selection, we ignore all punctuation marks, and other symbols not represented in the current layout. Thus, to select {\sl Thomas Stearns Eliot}, abbreviated in our list as {\sl T.S., Eliot}, a user must generate events for letters ``T'' (\emph{up}) and ``S''(\emph{left}), omitting punctuation marks and spaces. On the top, bottom, and sides of the list, we can see groups of letters displayed. This is a visual aid, showing the current layout. The groups are arranged in accordance with the direction in which a user must tilt the joystick to generate an event for this group. Some of these letters are grayed out. That means that in the current list, with given selection \emph{prefix}, there are no entries which contain this letter as the next choice. Graying them out helps a user to find the next letter more easily.  The {\sl AccelKey} selection method could be used with languages using alphabets besides the Latin. For example on Figure~\ref{fig:screen_cyr} we can see an example using the Cyrillic alphabet. 

\section{Further Work}

One direction for future research is the development of algorithms which would help to minimize the average number of events per list element, thereby optimizing layout.

Another interesting direction to explore is the use of touch-sensitive screens with this algorithm. It should be possible to use ``gestures'' performed by the user touching the screen as an input method.

On the implementation side, we would like to experiment further with different screen arrangements and visual and possibly audible feedback to users during selection process. A trackball jitter elimination algorithm could also be further fine tuned.

A more precise comparative evaluation of the alogrithm efficiency using Hicks-Hayman\cite{hicks,hyman} and Fitts\cite{mackenzie} laws is also in our plans.

\appendix

\section{Source Code}
\label{src}

\subsection{Fast4.hs}
\label{fast4hs}
\begin{small}
\begin{verbatim}
{-
  Fast4 Algorithm Implementation
  
  (c)2007 Vadim Zaliva <lord@crocodile.org>
-}

module Fast4(f4filter, eventSeq, countEvents) where
             
import List
import Maybe
    
-- | Definition of layout - a mapping of 4 directional 
-- events to group of characters
f4layout :: [[Char]]

{-
f4layout = ["qwertyuiop",
            "asdfg",
            "hjkl",
            "zxcvbnm"]
-}
f4layout = ["abcdefg", 
            "hijklmn", 
            "opqrstu", 
            "vwxyz"]
           
-- | A complete alphabet used in current layout
f4abc :: [Char]
f4abc = ((nub . concat) f4layout)

-- | Checks if given prefix (events sequence) match given sting
f4match :: [Int] -> String -> Bool
f4match [] _ = True
f4match _ [] = False
f4match (p:px) (s:sx) = if elem s f4abc then
                        (elem s (f4layout!!p)) && (f4match px sx)
                        else f4match (p:px) sx


-- | Filter list, removing all elements which do not match given prefix
f4filter :: [Int] -> [String] -> [String]
f4filter p l = filter (f4match p) l 


-- Followg code is used only for metric calculation --


-- | Given list of strings, will count number of events required to
-- select each of them from this list using Fast 4 Algorighm
countEvents :: [String] -> [Int]
countEvents l = map (\x -> eventSeq l [] x x) l

eventSeq :: [String] -> [Int] -> String -> String -> Int
eventSeq l p [] s = length p + scrollSeq l s
eventSeq l p (x:xs) s = min (length p + 1 + scrollSeq l s) (
                        if elem x f4abc then 
                           eventSeq (f4filter np l) np xs s
                        else eventSeq l p xs s)
                            where
                            np = p ++ [findEvent x]

-- | Returns number of events required to select item from the list in
-- "scroll" mode. Initially pointer is positioned in the middle of the
-- list
scrollSeq:: [String] -> String -> Int
scrollSeq l x = abs (fromMaybe 0 (findIndex (cmp x) l) - (div (length l) 2))
    where
    cmp a b = (intersect a f4abc) == (intersect b f4abc)

-- | Returns event number associated with given char in current layout
findEvent :: Char -> Int
findEvent x = fromMaybe 0 (findIndex (elem x) f4layout)



\end{verbatim}
\end{small}

\subsection{MultiTap.hs}
\begin{small}
\begin{verbatim}
{-
  Multitap Search Algorithm Implementation
  
  (c)2007 Vadim Zaliva <lord@crocodile.org>
-}

module MultiTap(countTaps, countTapAndScrolls) where
             
import List
import Maybe
import Scroll
    
-- | Definition of layout - a mapping of 4 directional 
-- events to group of characters
mtlayout :: [[Char]]

mtlayout = ["abc",
            "def",
            "ghi",
            "jkl",
            "mno",
            "pqrs",
            "tuv",
            "wxyz"
           ]
           
-- | A complete alphabet used in current layout
mtabc :: [Char]
mtabc = ((nub . concat) mtlayout)

-- | Given list of strings, will count number of events required to
-- select each of them from this list using MultiTap filtering
countTaps :: [String] -> [Int]
countTaps l = map (tapSeq lf) lf
               where 
               lf = map (intersect mtabc) l

-- | Given list of strings, will count number of events required to
-- select each of them from this list using MultiTap + scroll filtering
countTapAndScrolls :: [String] -> [Int]
countTapAndScrolls l = [tapCost x + tapSeq [is|(i:is)<-lf, i==x] xs | (x:xs)<-lf]
    where 
    lf = map (intersect mtabc) l

tapSeq :: [String] -> String -> Int
tapSeq _ [] = 0
tapSeq l (x:xs) = min (scrollDistance l (x:xs)) 
                  ((tapCost x) + (tapSeq [is|(i:is)<-l, i==x] xs))

tapCost:: Char -> Int
tapCost x = 1 + fromMaybe 0 (elemIndex x (fromMaybe "" (find (elem x) mtlayout)))






\end{verbatim}
\end{small}

\subsection{Scroll.hs}
\begin{small}
\begin{verbatim}
{-
  Scroll-to-select algorithms metrics caluclation
  
  (c)2007 Vadim Zaliva <lord@crocodile.org>
-}

module Scroll(countScrolls, scrollDistance) where
    
import Maybe
import List
         
countScrolls :: [String] -> [Int]
countScrolls l = map (scrollDistance l) l

-- | Returns number of events required to select item from the list in
-- "scroll" mode. Initially pointer is positioned in the beginning of the
-- list
scrollDistance :: [String] -> String -> Int    
scrollDistance l x = fromMaybe 0 (elemIndex x l)
\end{verbatim}
\end{small}

\end{document}